\documentclass[author,type1rest,genTeX]{nrc2}
\usepackage{amssymb}
\usepackage{bm}
\usepackage{graphicx}
\usepackage{cite}
\usepackage[french,english]{babel}
\volyear{73}{1995}%
\journalcode{cjp}%
\filenumber{\ }%
\begin{document}
%
%
\title{Optical phonons polarized along the \boldmath $c$ axis of YBa$_2$Cu$_3$O$_{6+x}$,
 for $ x$ = 0.5 $\rightarrow$ 0.95 \unboldmath }
%
%

\author{C.C.~Homes}
\present{Condensed Matter Physics and Materials Science Department, Brookhaven
 National Laboratory, Upton, NY 11973, USA}
\address{Department of Physics and Astronomy, McMaster University,
 Hamilton, ON  L8S 4M1, Canada.}
\author{T.~Timusk}
\correspond{timusk@mcmaster.ca}
\author{D.A.~Bonn, R.~Liang and W.N.~Hardy}
\address{Department of Physics and Astronomy, University of British Columbia,
Vancouver, BC, V6T 2A6 Canada.}
\shortauthor{Homes et al.}
\dedication{This article appeared in a special issue of the Canadian Journal of
Physics in honour of B.N. Brockhouse.}
%
%
\received{May 8, 1995}%
\accepted{July 14, 1995}%

%
%
%
%
%
\begin{abstract}
The {\it c}-axis polarized phonon spectra of single crystals of
YBa$_2$Cu$_3$O$_{6+x}$, were measured for the doping range $x=0.5\rightarrow
0.95$, between 10~K and 300~K.  The low background electronic conductivity,
determined by Kramers-Kronig analysis of the reflectance, leads to a rich
phonon structure.  With decreased doping the five normally-active $B_{1u}$
modes broaden and the high-frequency apical oxygen mode splits into two
components. We associate the higher of these with the two-fold coordinated
copper ``sticks''.  The 155 cm$^{-1}$ low-frequency mode, which involves the
apical and the chain-oxygens, splits into at least three components with
decreasing doping.  Some phonon anomalies which occur near $T_c$ in the
highly-doped material occur well above $T_c$ in the oxygen-reduced systems.  An
unusual broad phonon band develops in the normal state at $\approx 400$
cm$^{-1}$, which becomes more intense at low doping and low temperatures,
borrowing oscillator strength from apical and plane oxygen modes resulting in a
major transformation of the phonon spectrum below $\approx 150$~K.
\end{abstract}
%
%
\begin{resume}
Les spectres de phonons polaris\'{e}s suivant l'axe {\it c}, dans des
monocristaux de YBa$_2$Cu$_3$O$_{6+x}$ ont \'{e}t\'{e} measur\'{e}s, pour
l'intervalle de dopage $x=0,5$ \`{a} $0,95$, entre 10 et 300~K.  La faible
condutivit\'{e} \'{e}lectronique de fond, d\'{e}termin\'{e}e par analyse
Kramers-Kronig de la r\'{e}flectance, donne lieu \`{a} une riche structure de
phonons.  Lorsqu'on diminue le dopage, les cinque modes $B_{1u}$ normalement
actifs s'\'{e}largissent, et le mode apical \`{a} haute fr\'{e}quence de
l'oxyg\`{e}ne se divise en deux composantes.  Nous associons la plus haute de
ces composantes aux \og\ b\^{a}tons \fg\ de cuivre \`{a} coordinence deux. Le
mode de basse fr\'{e}quence, \`{a} 155~cm$^{-1}$, qui implique les oxyg\`{e}nes
apical et en cha\^{i}ne, se s\'{e}pare en au moins trois composantes lorsque le
dopage diminue.  Certaines anomalies de phonons, pr\'{e}sentes au voisinage de
$T_c$ dans les cristaux faiblement dop\'{e}s, apparaissent \`{a} des
temp\'{e}ratures bien plus grandes que $T_c$ dans les syst\`{e}mes avec moins
de l'\'oxyg\`{e}ne.  A environ 400~cm$^{-1}$, il se d\'{e}veloppe dans
l'\'{e}tat normal une bande large insolite qui devient plus intense \`{a}
faible dopage et \`{a} basse temp\'{e}rature, avec une force d'ocsillateur
emprunt\'{e}e aux modes apical et plan de l'oxyg\`{e}ne, ce quie entra\^{i}ne
une transformation majeure des spectres de phonons au-dessous d'environ 150~K.
\\
\ \\
{[Traduit par la r\'{e}daction]} \\
%
%
%
%
\PACS{63.20.Dj, 74.25.Gz, 74.72.Bk, 78.30.-j}
\end{resume}
\maketitle

%
%
\section{Introduction}
The role of phonons in high-temperature superconductors is complex. On the one
hand there is ample evidence that the cuprates are not conventional
electron-phonon BCS superconductors. This evidence includes, for the optimally
doped materials a linear resistivity, in the normal state, with a
zero-tem\-pera\-ture intercept at zero resistance \cite{Gurvitch87,Martin90}, a
collapse of transport scattering at the superconducting transition
\cite{Romero92,Bonn92b,Yu92}, and a lack of an isotope effect
\cite{Batlogg87,Bourne87,Franck90,Crawford90b}.

On the other hand, a large number of studies show that certain phonons at least
are coupled to the carriers, and that this coupling is unconventional
\cite{Litvinchuk94,Pintschovius94}. The frequencies and widths of certain
phonon lines, again in the optimally-doped materials (precisely those where the
isotope effect is small), undergo abrupt changes at $T_c$.  This effect was
first observed by infrared spectroscopy
\cite{Bonn87b,Wittlin87a,Litvinchuk92,Macfarlane87}, then by Raman spectroscopy
\cite{Macfarlane87, Friedel90b,Altendorf93}, resonant neutron-absorption
spectroscopy \cite{Mook90} and neutron scattering \cite{Pyka93}.

Broad minima in the {\it ab}-plane optical conductivity are seen at frequencies
that are in the phonon region. These have been shown to be associated with {\it
c}-axis longitudinal optic (LO) phonons \cite{Timusk91,Reedyk92e} that couple
to the {\it ab}-plane conductivity by a symmetry-breaking mechanism.

In this paper we focus on the changes in the {\it c}-axis phonon spectra of
YBa$_2$Cu$_3$O$_{6+x}$ as we change the oxygen doping.  These spectra can be
separated from the weak and smooth electronic background conductivity whose
intensity rises with doping from a few $\Omega^{-1}{\rm cm}^{-1}$ in  the
$x=0.5$ material\footnote{C.C. Homes, T. Timusk, D.A. Bonn, R. Liang and W.N.
Hardy. Unpublished results.} to 450~$\Omega^{-1}{\rm cm}^{-1}$ in overdoped
samples \cite{schutzmann94}. The background is due to inter-cell hopping of
free carriers \cite{Cooper94}. Superimposed on this electronic background are
five strong phonon lines. These five lines are also seen the in the spectra of
ceramic superconductors \cite{bonn88a,genzel89}, and have been studied by a
large number of groups.  The extensive literature on ceramic spectra has been
summarized by Feile in ref.~\citen{feile89}. However, in ceramics, effects due
to the background {\it ab}-plane conductivity distort the spectra, particularly
the line strengths, and the {\it c}-axis polarized phonons are best studied in
single crystals. Single crystal studies are difficult due to the small area of
the available flux-grown samples.  Nevertheless, a number of studies have
appeared, with YBa$_2$Cu$_3$O$_{6+x}$ being the most studied material [see
footnote 3 and
refs.~\citen{schutzmann94,bozovic87,collins89c,koch90,koch90b,cooper92,homes93a}].
In the present work single crystals with cleaved surfaces are used and the
region of doping ranges from the optimally doped, that is the material with the
highest $T_c=93.5$~K with $x=0.95$, to an underdoped sample with $x=0.5$ which
is still superconducting with a $T_c=53$ K. Preliminary reports of this work
have been published \cite{homes93a, timusk92b,timusk95a}, and the electronic
background is the subject of a separate publication [see footnote 3].

\section{Experimental Details}
\subsection{Experimental}
The spectra were obtained by a Kramers-Kronig transformation of {\it c}-axis
polarized reflectance of millimeter-size, high-quality single crystals. The
growth of the single crystals and experimental techniques used have been
reported on previously \cite{homes93a,liang92}. To maximize the signal from the
{\it ac} face of the crystal, an infrared beam larger than the crystal was
used. The sample was coated with gold {\it in situ} and then remeasured at each
temperature to obtain a reference spectrum \cite{homes93b}.

\begin{figure}[t]
\caption{The reflectance of YBa$_2$Cu$_3$O$_{6+x}$ for radiation polarized
along the {\it c}-axis from $\approx 50$ to 700 cm$^{-1}$ at several
temperatures above and below $T_c$, for two oxygen dopings $x$: ({\it a})
fully-doped material with $x=0.95$ ($T_c=93$~K), and ({\it b}) an underdoped
crystal with $x=0.6$ ($T_c=58$~K).  The main change in the phonon spectrum in
the underdoped material is a new line at 610~cm$^{-1}$ that we associate with
the bridging oxygen at the two-fold coordinated copper sites in the chain
layer. Another new feature in the $x=0.6$ material is a broad band at $\approx
400$ cm$^{-1}$ that appears at low temperature.}%
\vspace*{0.4cm}%
\centerline{\includegraphics[width=11.0cm]{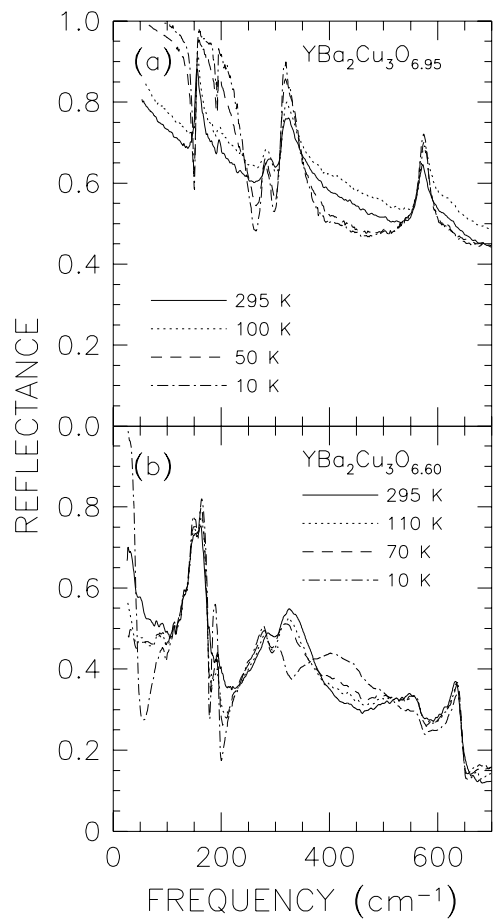}}%
\vspace*{1.2cm}%
\label{reflec}%
\end{figure}

The reflectance of YBa$_2$Cu$_3$O$_{6+x}$, for two oxygen concentrations,
$x=0.6$ and 0.95, for light polarized parallel to the {\it c} axis is shown in
the phonon frequency range in Fig.~\ref{reflec} at several temperatures. The
reflectance has a continuous background, rising towards unity at low frequency,
characteristic of a poor metal, with sharp resonances due to optical phonons.
In the superconducting state, the development of the superconducting condensate
has the effect of producing a reflectance edge which shifts down to lower
frequency as the doping is reduced. Also, as the temperature is lowered, the
shapes of the phonon features change due to the changing background
conductivity.

Changes in oxygen doping have two principal effects on the phonon spectra:
first, as the doping level is lowered, a new phonon feature appears at $\approx
610$ cm$^{-1}$. Secondly, a very broad feature, weak at first, but stronger at
low oxygen dopings, appears in the normal state at $\approx 400$ cm$^{-1}$, and
becomes stronger at low temperature. These effects are seen more clearly in the
optical conductivity, discussed in the following section.

\subsection{Optical conductivity}
A Kramers-Kronig transformation of the reflectance was us\-ed to calculate the
optical conductivity. The reflectance was extended to regions outside the
actual measurements as follows: below the lowest measured frequency
Hagen-Rubens frequency dependence: [$(1-R)\propto \omega^{-1/2}$] was assumed
to hold in the normal state.  In the superconducting state, where a plasma edge
develops, it was assumed that the reflectance was given by
$(1-R)\propto\omega^{-2}$. This assumes that the conductivity is finite and
constant to zero frequency. There is good evidence from comparisons with
transport measurements that both assumptions are valid [see footnote 3].
Temperature-dependent reflectance measurements were carried out to
5000~cm$^{-1}$. The room-temperature measurements of Koch et al. \cite{koch90}
were used from 5000 cm$^{-1}$ to $3.5\times 10^5$ cm$^{-1}$ and above this
frequency free-electron behavior was assumed: $R\propto \omega^{-4}$.

%
%

\begin{figure}
\caption{The optical conductivity ($\sigma_1$) at 295~K for
YBa$_2$Cu$_3$O$_{6+x}$ for radiation polarized along the {\it c}-axis from
$\approx 50$ to 700 cm$^{-1}$ for five oxygen dopings, $x=0.5 \rightarrow
0.95$. As the doping is reduced, the background conductivity is reduced
uniformly at all frequencies.  At high frequency, in addition to the apical
oxygen [O(4)] peak at 570~cm$^{-1}$ a new phonon peak grows at $\approx
610$~cm$^{-1}$; this peak is associated with the O(4) mode and the growing
density of two-fold coordinated copper sites in the underdoped materials.  As
the doping is reduced, the O(4) modes acquire an increasingly dispersive line
shape.}
\vspace*{-0.4cm}%
\centerline{\includegraphics[width=10.5cm]{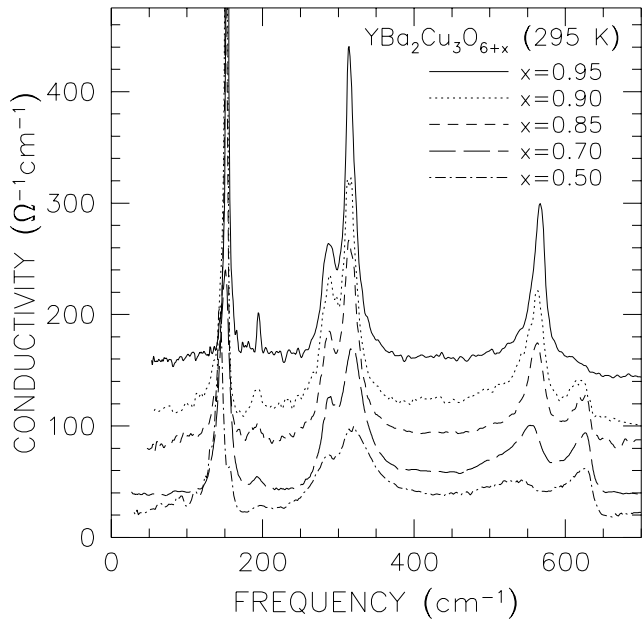}}%
\vspace*{-1.5cm}%
\label{sigma295}
\end{figure}

The 295~K conductivity along the {\it c}-axis is shown in Fig.~\ref{sigma295}
for a wide range of dopings. The 10~K conductivity is shown in
Fig.~\ref{sigma10}. Superimposed on a continuous background that increases in
strength with doping are a series of sharp phonon lines. In general, the
strength of the phonon bands does not depend strongly on doping, but as the
doping is reduced a new phonon appears  at $\approx 610$~cm$^{-1}$ in the
nearly fully-doped material and moves to higher frequency as the oxygen
concentration is reduced. We will call this the 610 cm$^{-1}$ line in what
follows, but its actual frequency changes with doping and is as high as
635~cm$^{-1}$ in the $x=0.5$ material. The other major effect of doping occurs
at low temperature where a new broad band appears at 400~cm$^{-1}$ as the
doping level is reduced to $x=0.7$. This band seems to increase its oscillator
strength at the expense of the two high-frequency modes and the band at
310~cm$^{-1}$. Also, its center frequency appears to decrease as the doping is
reduced, reaching a value of $\approx 375$~cm$^{-1}$ in the $x=0.5$ crystal.

%
%

\section{Results}
%
%
\subsection{Calculation of the phonon parameters}
In what follows we refer to the six phonons that we observe by their room
temperature frequencies in the fully-doped material, bearing in mind that the
actual center frequencies change with both temperature and doping. There are
five strong phonons, and at room temperature their frequencies are: 155, 194,
279, 312 and 570 cm$^{-1}$, plus a weaker one at 610 cm$^{-1}$.  All the lines
have symmetric line shapes except for the two highest ones. The 570 cm$^{-1}$
and the 610 cm$^{-1}$ lines have Lorentzian shapes at room temperature, but at
low temperature they become asymmetric \cite{burlakov}.  Such lines are usually
described in terms of a Fano line shape
\begin{displaymath}
  \sigma_1(\omega) = A \left[ {{(x+q)^2}\over{(1+x^2)}} \right]
\end{displaymath}
where $\sigma_1(\omega)$ is the conductivity, $A$ is a constant, $x
 =(\omega - \omega_i)/\gamma_i$, $\omega_i$ is the phonon frequency, $\gamma_i$ is
the phonon line width and $q_i$ is a parameter that describes the asymmetry of
the $i$th phonon \cite{fano61}.

%
%
\begin{figure}
\caption{The {\it c}-axis optical conductivity ($\sigma_1$) at 10~K for five
oxygen dopings, $x = 0.5 \rightarrow 0.95$. The curves have been offset
vertically in 200 $\Omega^{-1}$cm$^{-1}$ increments for clarity.  There is a
major redistribution of spectral weight from the O(4) peaks at $\approx 570$
cm$^{-1}$ and $\approx 610$~cm$^{-1}$, and the in-plane buckling mode at 310
cm$^{-1}$, to an new broad band in the $\approx 400$ cm$^{-1}$ range.  The peak
shifts to lower frequency as the doping is reduced, starting as a shoulder on
the apical-oxygen line at 550~cm$^{-1}$ and ending up centered at $\approx 375$
cm$^{-1}$ in the $x=0.5$ sample.}
\vspace*{0.4cm}%
  \centerline{\includegraphics[width=10.5cm]{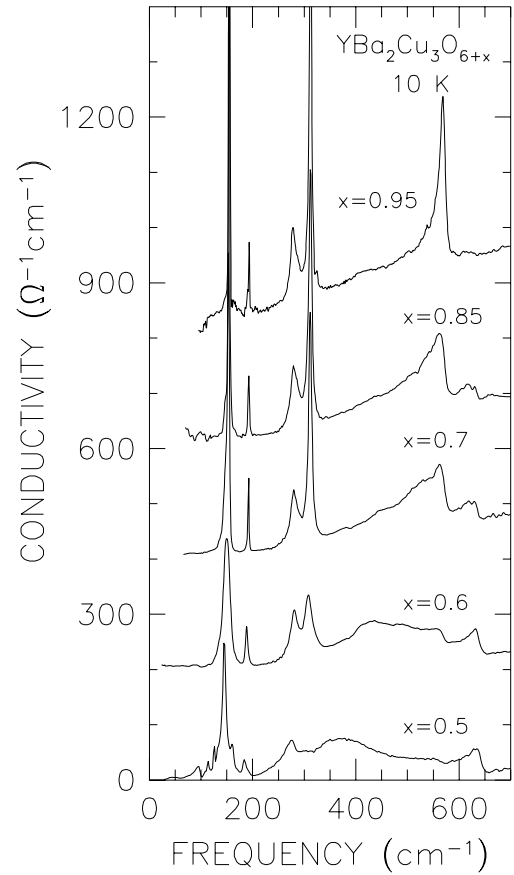}}%
\vspace*{1.2cm}%
\label{sigma10}
\end{figure}

%
%
\begin{table*}[t]
\topcaption{The parameters used to fit Lorentzian line shapes to the
 peaks in the optical conductivity of YBa$_2$Cu$_3$O$_{6+x}$, for
 $x=0.95, 0.85$, $0.7$, and $0.5$ at 10~K.  The two high-frequency
 copper-oxygen vibrations have been fitted using non-zero rotations.$^a$ }
\begin{tabular}{ccccccccccccccc}
\hline*
 \multicolumn{3}{c}{$x=0.95$} & & \multicolumn{3}{c}{$x=0.85$} & &
 \multicolumn{3}{c}{$x=0.7$}  & & \multicolumn{3}{c}{$x=0.5$} \\
 \cline{1-3} \cline{5-7} \cline{9-11} \cline{13-15}
 $\omega_{{\rm TO},i}$ & $\gamma_i$ & $\omega_{pi}$ &  & $\omega_{{\rm TO},i}$ &
 $\gamma_i$ & $\omega_{pi}$ & & $\omega_{{\rm TO},i}$ & $\gamma_i$ &
 $\omega_{pi}$ & & $\omega_{{\rm TO},i}$ & $\gamma_i$ & $\omega_{pi}$ \\
\hline
%
%
 155 & 2.0 & 398 &
 $\left\{ \begin{array}{c}  \\ \end{array} \right.$ \hspace*{-0.8cm}&
 \begin{tabular}{c} 147 \\ 153 \end{tabular} &
 \begin{tabular}{c}  8.4 \\ 3.8 \end{tabular} &
 \begin{tabular}{c} 248 \\ 351 \end{tabular} &
 $\left\{ \begin{array}{c}  \\ \\     \end{array} \right.$ \hspace*{-0.8cm} &
 \begin{tabular}{c} 136 \\ 147 \\ 153 \end{tabular} &
 \begin{tabular}{c}  15 \\ 9.8 \\ 4.2 \end{tabular} &
 \begin{tabular}{c} 167 \\ 322 \\ 267 \end{tabular} &
 $\left\{ \begin{array}{c}  \\ \\ \\ \\    \end{array} \right.$ \hspace*{-0.8cm} &
 \begin{tabular}{c}  93 \\ 113 \\ 125 \\ 145 \\ 161 \end{tabular} &
 \begin{tabular}{c}  11 \\ 5.0 \\ 5.2 \\ 9.0 \\ 5.9 \end{tabular} &
 \begin{tabular}{c} 116 \\  80 \\ 117 \\ 363 \\ 120 \end{tabular} \\
%
%
  194 & 2.2 & 128 & $\rightarrow$ \hspace*{-0.8cm} &
  192 & 2.8 & 133 & $\rightarrow$ \hspace*{-0.8cm} &
  189 & 4.2 & 133 & $\rightarrow$ \hspace*{-0.8cm} &
  184 & 5.9 & 110 \\
%
%
  279 & 19  & 357 & $\rightarrow$ \hspace*{-0.8cm} &
  281 & 19  & 329 & $\rightarrow$ \hspace*{-0.8cm} &
  280 & 22  & 323 & $\rightarrow$ \hspace*{-0.8cm} &
  272 & 33  & 299 \\
  312 & 5.2 & 487 & $\rightarrow$ \hspace*{-0.8cm} &
  312 & 7.0 & 417 & $\rightarrow$ \hspace*{-0.8cm} &
  309 & 18  & 333 & $\rightarrow$ \hspace*{-0.8cm} &
  301 & 28  & 184 \\
%
%
  570 & 14 & 483 &
 $\left\{ \begin{array}{c} \\  \end{array} \right.$ \hspace*{-0.8cm} &
 \begin{tabular}{c} 566 \\ 619 \end{tabular} &
 \begin{tabular}{c}  23 \\  23 \end{tabular} &
 \begin{tabular}{c} 321 \\ 205 \end{tabular} &
 $\left\{ \begin{array}{c} \\  \end{array} \right.$ \hspace*{-0.8cm} &
 \begin{tabular}{c} 564 \\ 630 \end{tabular} &
 \begin{tabular}{c}  30 \\  24 \end{tabular} &
 \begin{tabular}{c} 188 \\ 250 \end{tabular} &
 $\left\{ \begin{array}{c} \\  \end{array} \right.$ \hspace*{-0.8cm} &
 \begin{tabular}{c} 551 \\ 637 \end{tabular} &
 \begin{tabular}{c}  30 \\  24 \end{tabular} &
 \begin{tabular}{c} 127 \\ 251 \end{tabular} \\
\hline*
\end{tabular} \\
{\small
\indent  NOTE: All parameters are in cm$^{-1}$. \\
\ $^a$ The two high-frequency apical-oxygen modes have
 been fit using a Lorentzian with a phase.  The angles (in radians)
 used are for $x=0.95$, $\theta=-0.40, 0.0$, for $x=0.85$, $\theta=-0.50,
 -0.15$, for $x=0.7$, $\theta=-0.62, -0.45$, for $x=0.5$, $\theta=-0.75,
 -0.80$, for the low- and high-frequency components, respectively.
}
\vspace*{0.3cm}%
\label{doping}%
\end{table*}

While the Fano line shape describes the 570~cm$^{-1}$ phonon at low
temperatures, it fails as the line shape becomes more Lorentzian
($q\rightarrow\infty$).  A better fit to the data is in terms of a classical
Drude-Lorentz model for the dielectric function, for each phonon band, but with
an empirical line shape made up of a mixture of the real and imaginary parts of
the Lorentzian oscillator:
\begin{displaymath}
  \tilde\epsilon(\omega) = { {\omega_p^2\,e^{i\theta}} \over {\omega_{\rm TO}^2
  - \omega^2 - i\omega\gamma}} \hspace*{4.0cm}  (1)
\end{displaymath}
where $\omega_{p}$ is the effective plasma frequency, $\omega_{\rm TO}$ the
center frequency and $\gamma$ the width of the phonon. The parameter $\theta$
is the phase associated with the asymmetry of the phonon. The line shape is
that of a classical oscillator for $\theta=0$ but becomes asymmetric for
nonzero $\theta$.

An advantage of this rotated-Lorentzian line shape over the Fano profile is
that one can associate with it an oscillator strength.  For a pure Lorentzian,
the integrated oscillator strength is defined to be $\omega_{p}^2/8$; for the
rotated Lorentzian, to leading order the oscillator strength is
$\cos(\theta)\omega_p^2/8$, so that for small rotations, the introduction of an
asymmetric line shape does not alter the oscillator sum rule too much. However,
for rotations of greater than $\approx 0.4$ radians, the error introduced is
greater than 10\%, and the other contributions to the sum rule begin to play an
increasingly large role, further compounding this error.  The asymmetry of the
rotated Lorentzian is related to the Fano parameter $q$ by $q^{-1} \propto\tan
(\theta/2)$.

All the phonon lines have been fitted to (1) with a non-linear least-squares
method.  In addition to the parameters of the modified-Lorentzian oscillator, a
linear background was used in the fits.  We find that within the accuracy of
our experiments, the four low-frequency lines are symmetric and only the
570~cm$^{-1}$ line, and the new feature associated with it at 610~cm$^{-1}$,
have clearly asymmetric shapes which become more pronounced at low
temperatures.

It should be noted that if the Kramers-Kronig analysis is done on data that
deviates from the true reflectance, for example by using room-temperature data
for the high-frequency extrapolation for a low-temperature data set, there is a
tendency for sharp lines to acquire an artificial asymmetry.  As the
experimental data become more accurate, the lines become more symmetric.  The
lines that have the highest reflectance are most susceptible to this error,
which in our case is the 155~cm$^{-1}$ line, which we find accurately
symmetric, thus confirming the position of our 100\% line. We find that an
error of greater than 2\% in the unity reflectance line would show an
observable asymmetry.

The phonon parameters obtained by the least-squares fitting procedure applied
to the rotated-Lorentzian profile with of all the measured spectra are
assembled in Table~\ref{doping}. We estimate the error in the parameters to be
$\approx 1$\% for the center frequencies, $\approx 5$\% for the line widths,
and $\approx 5$\% for the effective plasma frequencies.  Where there is a
finite rotation angle $\theta$, the error could be larger.

%
%
%
%
\subsection{Assignment of phonons}
\begin{figure}[t]
\caption{The orthorhombic unit cell of YBa$_2$Cu$_3$O$_7$.  The chain oxygen is
denoted as O(1), the apical copper as Cu(1), and the O(4) as the apical oxygen
atom.  When all the O(1) sites are occupied, the Cu(1) atom is four-fold
coordinated.  Deoxygenation results in the oxygens at the O(1) sites being
removed, lowering the coordination of the Cu(1) atom.}
\vspace*{0.4cm}%
\centerline{\includegraphics[width=8.0cm]{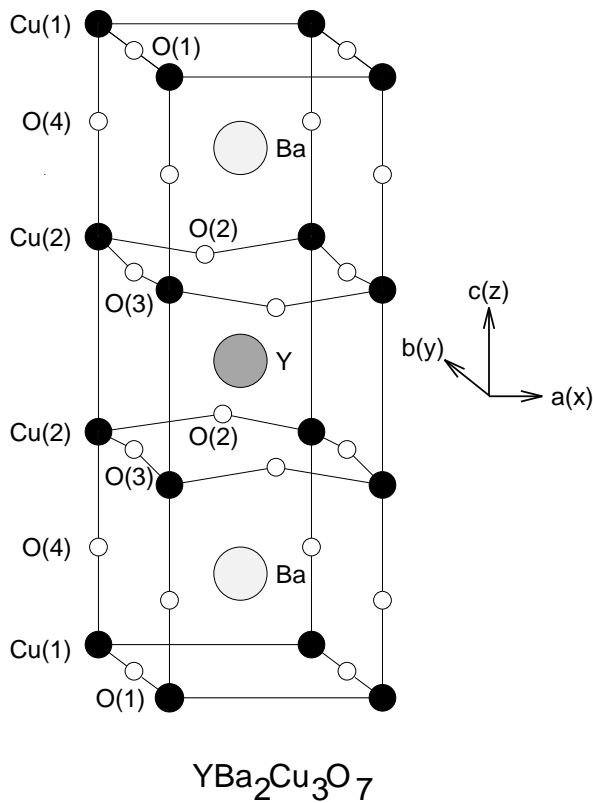}}%
\vspace*{0.2cm}%
\label{struct}
\end{figure}

Group theory predicts seven $B_{1u}$ infrared-active modes polarized in the
c-direction for the fully-oxygenated ortho\-rhombic (Ortho~I)
YBa$_2$Cu$_3$O$_7$ shown in Fig.~\ref{struct} \cite{bates87,kress88}. There is
crystallographic evidence that the oxygen-reduced material, with a plateau in
the $T_c$ vs $x$ graph (in the $T_c\approx 60$~K region), has a doubled unit
cell normal to the chain direction, resulting from an alternation of
fully-occupied O(1) sites on chains with empty sites. This structure (Ortho~II)
is expected to have 13 infrared-active modes, but many of the new modes result
from extremely small splittings of modes involving atoms far away from the
chains. These small splittings may be partially responsible for the systematic
broadening of most of the phonon lines at reduced dopings.  We expect the
largest changes for modes that involve atoms on and near the chains.

%
%

%
%

\begin{table}
\caption{The fitted phonon parameters for the {\it c}-axis $B_{1u}$ modes in
YBa$_2$Cu$_3$O$_{6.95}$ at 295~K determined from optical and inelastic
neutron-scattering measurements. The effective-plasma frequency based on the
shell model with formal charges in column 5 [Y = +3.0, Ba = +2.0, Cu1 = Cu2 =
2, O(1)
 = O(2) = O(3) = O(4) = -2] and fitted charges in column 6, [Y = +2.8, Ba = +2.5,
Cu1 = Cu2 = 2, O(1)= -1.8, O(2) = -1.8, O(3) = -.17, O(4) = -2.6].}
\begin{tabular}{ccccccc}%
\hline*
 \multicolumn{3}{c}{Optical$^{\rm a}$} & &
 \multicolumn{3}{c}{Neutron$^{\rm b}$} \\
\cline{1-3} \cline{5-7}
 $\omega_{{\rm TO}i}$ & $\gamma_i$ & $\omega_{pi}$ & & $\omega_i$ &
 $\omega_{pi}^{\rm c}$ & $\omega_{pi}^{\rm d}$ \\
\hline
 ---             &  --- & --- & & 111 & 100 & 115 \\
 153             &  3.6 & 389 & & 149 & 283 & 330 \\
 ---             &  --- & --- & & 186 &  22 & 14  \\
 195             &  3.7 &  95 & & 198 &  79 & 94  \\
 287             & 18.3 & 316 & & 291 & 455 & 400 \\
 315             & 15.6 & 498 & & 322 & 493 & 457 \\
 \ \ \ 568$^{\rm e}$\ \ \  & \ \ \ 18.4\ \ \  & \ \ \ 408\ \ \  & \ \ &
 \ \ \ 556\ \ \  & \ \ \ 256\ \ \  & \ \ \ 365\ \ \  \\
\hline*
\end{tabular} \\
\small{
 NOTE: All units are in cm$^{-1}$. \\
 $^a$Fitted to optical conductivity, this work. \\
 $^b$W.~Reichardt et al. Private communication. \\
 $^c$Effective charges, formal valence charges. \\
 $^d$Effective charges, fitted to optical data. \\
 $^e$The high-frequency apical-oxygen mode has been fitted using an asymmetric
line shape described by a Lorentzian with a rotation angle of $\theta=-0.27$
rad.
}
\vspace*{0.3cm}
\label{neutrons}
\end{table}

The assignment of the infrared-active spectral lines to normal modes is most
reliable when the results from inelastic-neutron scattering are combined with a
realistic lattice-dynamic model \cite{cowley63}. Such a procedure has been
carried out recently for fully-oxygenated YBa$_2$Cu$_3$O$_7$,\footnote{W.
Reichardt. Private communication.} and we make extensive use of the
eigenvectors of these models in our discussion of {\it c}-axis the phonon
spectra. A list of calculated contributions the various modes to the
polarizability has been published by Timusk et al. \cite{timusk95a}.  It is
clear from the shell models that infrared modes, unlike Raman modes, which can
often be assigned to the motion of a particular ion, usually involve the motion
of all the ions in the unit cell.  The shell-model eigenvectors suggest that
there are some exceptions.  For example the 570 cm$^{-1}$ mode is almost a pure
mode of the apical O(4) oxygen (a description of the different copper and
oxygen environments is shown in Fig.~\ref{struct}).  Similarly, the 279
cm$^{-1}$ mode is an O(1) chain-oxygen vibration, polarized in the {\it
c}-direction and the 315 cm$^{-1}$ mode an O(2) and O(3) plane-bending
vibration. The low-frequency modes involve many more atoms, and while the 194
cm$^{-1}$ mode has been identified as the yttrium mode, it also involves
substantial motion of the O(2) planar-oxygen atoms. The 155 cm$^{-1}$ mode,
often called the barium mode, according to the shell model gets more spectral
weight from the O(4) oxygen motion than from the barium, and also involves the
motions of the Cu(1) and O(1) chain-copper and -oxygen atoms.  This prediction
is in agreement with our observation that this modes splits in oxygen-reduced
materials.

%
%
\begin{figure*}[t]
\caption{The variation of the central frequency ($\omega_{{\rm TO},i}$), width
($\gamma_i$) and the effective plasma frequency ($\omega_{pi}$) as a function
of doping ($x=0.5\rightarrow 0.95$) of the five strong phonons observed in the
highly-doped materials, at 10~K ($\bullet$).  The room-temperature frequencies
have been included for the three central panels ($\ast$).  Also included in the
panels are two phonons that appear only at lower oxygen dopings ($\circ$): in
the first panel, at new mode appears at $\approx 148$ cm$^{-1}$ and gains
oscillator strength as the mode observed in the highly-doped materials at
$\approx 155$ cm$^{-1}$ becomes weaker.  The second phonon appears at $\approx
610$ cm$^{-1}$ and, similarly to the 148 cm$^{-1}$ phonons, gains oscillator
strength with decreasing doping.  Inset: rotation angles for the $\approx 570$
cm$^{-1}$ mode ($\bullet$), and the $\approx 610$ cm$^{-1}$ mode ($\circ$).}
\vspace*{-6.8cm}%
\centerline{\includegraphics[width=16.3cm]{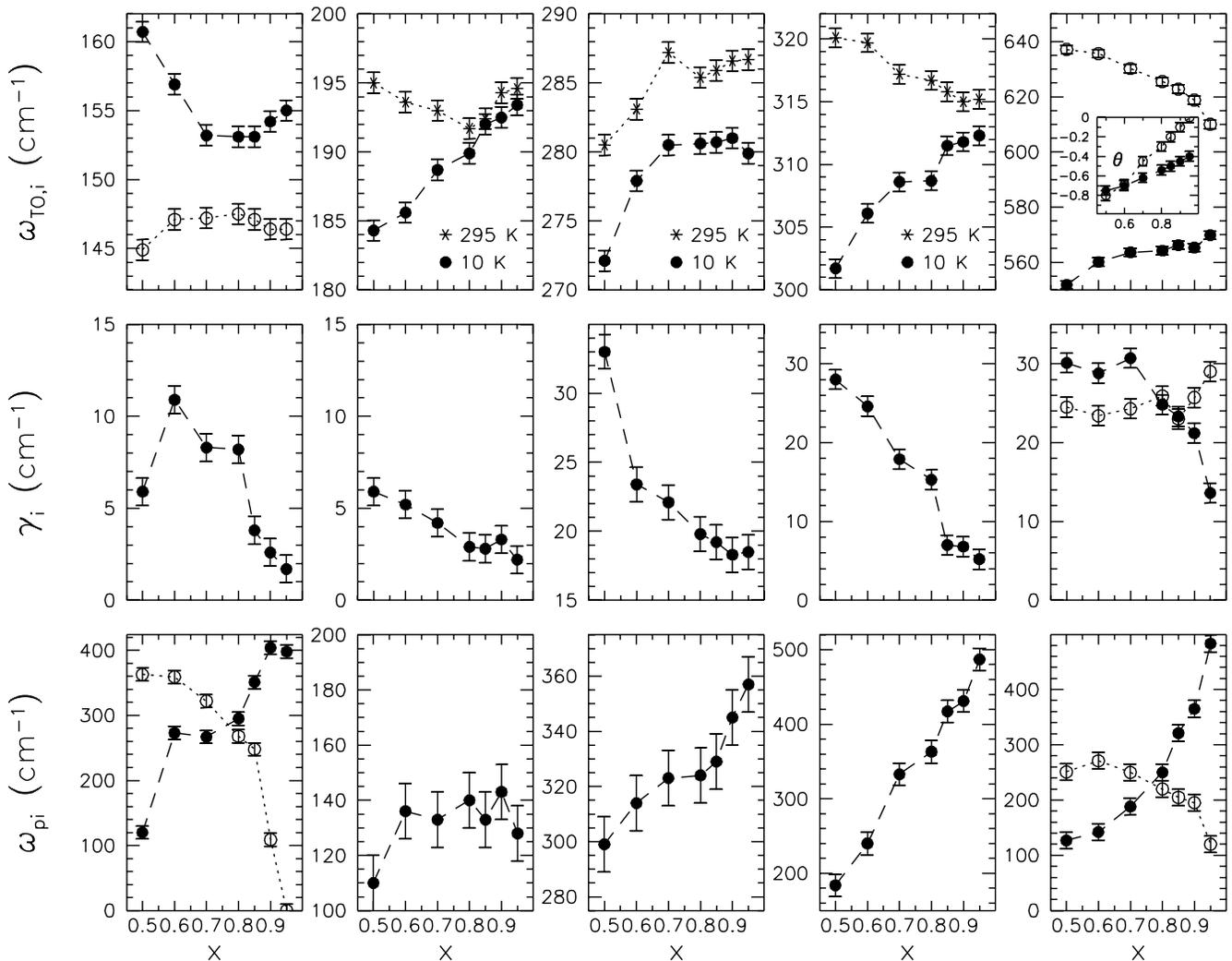}}%
\vspace*{-0.5cm}%
\label{parms}
\end{figure*}

%
%
\begin{figure}
\caption{The evolution of the shape of the 155 cm$^{-1}$ line, at 295~K and
10~K, for five oxygen dopings ($x=0.5 \rightarrow 0.95$).  As the oxygen doping
is reduced, the line broadens, first developing shoulders, and then finally
splitting into several clearly resolved components.}
\vspace*{0.3cm}%
\centerline{\includegraphics[width=10.5cm]{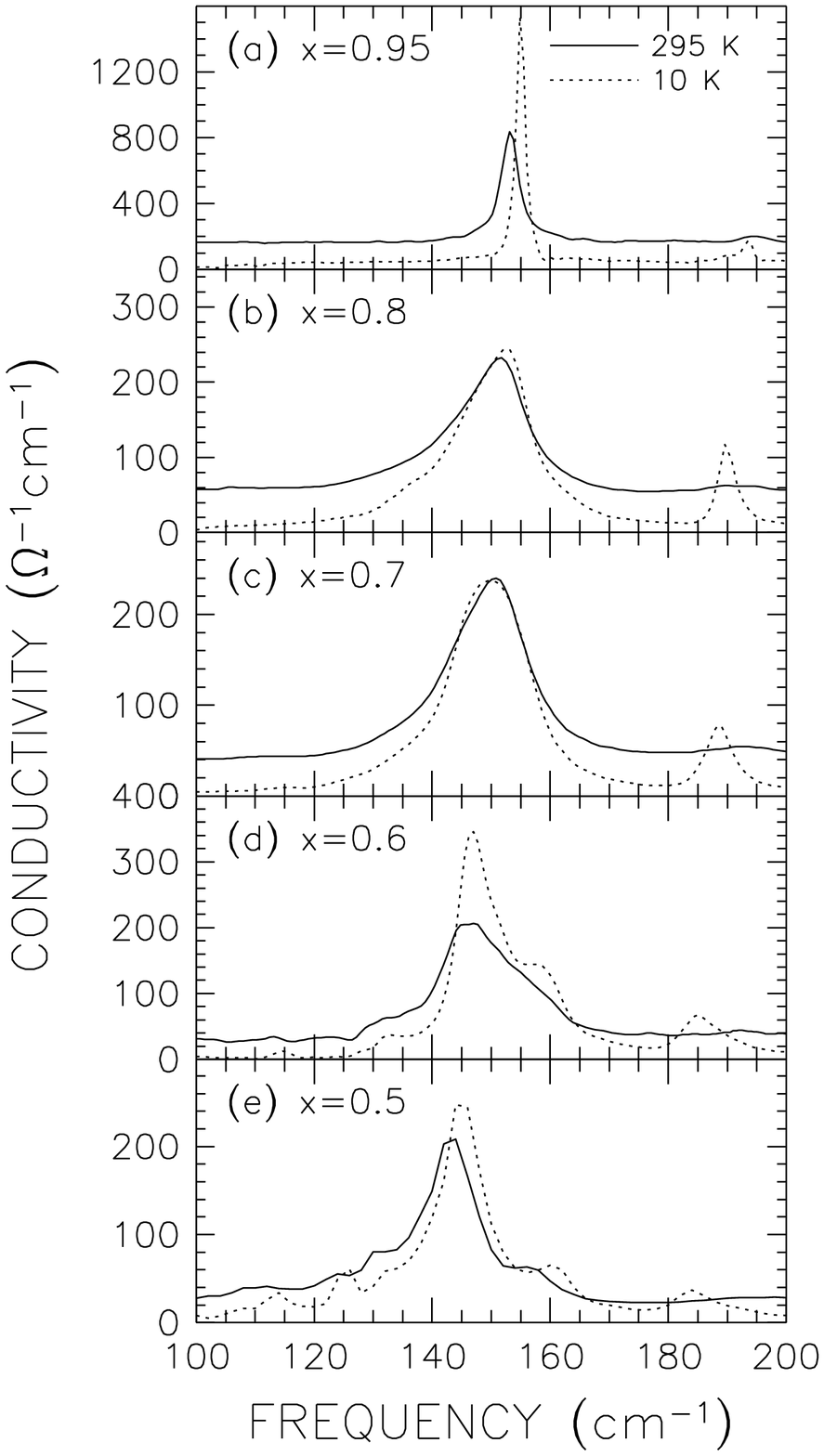}}%
\vspace*{1.3cm}%
\label{barium}
\end{figure}

In the highly-doped YBa$_2$Cu$_3$O$_{6.95}$ material five strong pho\-nons are
observed at 155, 194, 279, 312 and 570~cm$^{-1}$ at 10~K. These frequencies are
in excellent agreement with previous measurements on ceramics
\cite{bonn88a,Crawford88}.  Table~\ref{neutrons} compares these frequencies
with those of the neutron-based shell model. As expected, since the \boldmath
$k=0$ \unboldmath frequencies have been fitted to the model, the agreement is
satisfactory.  What is surprising is the excellent agreement between the
measured line strengths, and those predicted by the shell-model eigenvectors.
In calculating the line strengths we have used two separate approaches: the
simplest is to use the formal charges of the ions in all cases. The only case
where this seems to be inappropriate is for the 570~cm$^{-1}$ mode where the
observed line strength is larger, and a larger effective charge on the O(4)
would give better agreement between the model and the infrared observations.
This line is also anomalous in that it develops a strong asymmetry in the
oxygen-reduced materials, a sign of strong electron-phonon interaction which
may enhance the polarizability of the mode by dynamic charge transfer from the
O(4) to the planes.

The second approach is to use a least-squares fit to the effective charges to
the observed infrared oscillator strengths.  The results are shown in column
six of Table~\ref{neutrons}. The fitted results, as expected, agree better with
the optical data.  However, it is interesting to note that in order to get this
better fit, the main change was the removal of charge from the O(4) ion, and
the addition of that charge on the planar oxygen ions.

%
%

The two missing lines are predicted by the shell model to be very weak.  Of
these there is some evidence for the lower one predicted to be at 111 cm$^{-1}$
in some of our spectra, but they are near the limit of the accuracy of the
experiments.

Two modes not predicted by the shell model are seen: the broad band at $\approx
400$ cm$^{-1}$, and weak peak at $\approx 610$~cm$^{-1}$. In what follows we
will argue that the 610~cm$^{-1}$ line is associated with the two-fold
coordinated copper neighbor of the O(4) oxygen. Such sites are found in
oxygen-reduced material where the 610 cm$^{-1}$ line is much stronger, and its
presence in our $x=0.95$ sample is an indication that this sample contains a
small concentration of broken chain fragments with two-fold coordinated copper.

Overall, at room temperature at least, there is good agreement with the shell
model and the observed phonon spectrum and there is no need to invoke special
electronic enhancements to explain the intensities of the phonon lines, nor is
there any evidence of strong screening by the electronic background.

%
%
\subsection{Doping dependence of the phonon parameters}
As the doping level is reduced, a number of changes in the phonon parameters
take place; these are summarized for the five fundamental phonons observed in
the highly-doped material, as well as the two strong splittings associated with
the 155 and 570 cm$^{-1}$ modes at 10~K observed at lower dopings, in
Fig.~\ref{parms}.  To a first approximation one expects the phonon frequencies
to soften with the expansion of the {\it c}-axis lattice parameter. One
generally expects the frequency to vary approximately as the third power of the
lattice parameter \cite{Maroni89}.  Since the {\it c}-axis lattice parameter
increases by 0.5\% between $x=0.93$ to $x=0.50$, one would expect an overall
softening of the order of 1.5\% as the result of the lattice expansion

In very rough agreement with this simple picture, many mod\-es soften, but by a
larger amount of 3.5\% in this doping range. There are also several notable
exceptions. First there is a tendency, at room temperature, for several modes
to harden as the lattice expands on removal of oxygen, for example the 312
cm$^{-1}$ mode.  The 155 cm$^{-1}$ mode as well as the high-frequency O(4) mode
behave in a similar fashion: they split into two components, the higher
frequency component hardens, the lower frequency one softens.  As we will show
below for the 570 cm$^{-1}$ mode, and its high-frequency partner the 610
cm$^{-1}$ mode the higher frequency component corresponds to the O(4) at the
two-fold coordinated copper sites whereas the lower frequency corresponds to
the O(4) at the four-fold coordinated copper sites. While the shell model
eigenvectors suggest that the 155~cm$^{-1}$ mode also involves a large amount
of O(4) motion, many other atoms in the unit cell participate in the motion,
and the net result is a behavior that is opposite to that for the 570 cm$^{-1}$
mode: it is the high-frequency, hard component that loses oscillator strength.

The mode at $\approx 155$ cm$^{-1}$ in the highly-doped material has split into
a doublet in the $x=0.85$ material, and consists of at least three lines in the
$x=0.7$ material; the phonon lines are significantly broader than the narrow
line in the $x=0.95$ material.  Fig.~\ref{barium} shows the evolution of the
shape of this line with doping.  The multi-component nature of the line and the
gradual shift, with reduced doping to lower frequency components is reminiscent
of the 500~cm$^{-1}$ Raman line which has been shown to have many components,
associated with chain disorder in underdoped samples \cite{Hadjiev93}.  The
splitting of the $\approx 155$~cm$^{-1}$ mode is also associated with the loss
of O(1) chain oxygens.  Unlike the 570~cm$^{-1}$ mode, where the O(4)--Cu(1)
chain-copper bond is stretched, the 155~cm$^{-1}$ mode involves the O(4)--Cu(2)
bond which is also expected to change its length as the Cu(1) changes from
four- to two-fold coordination.

%
%
\begin{figure*}[t]
\caption{The temperature dependence of the frequency ($\omega_{{\rm TO},i}$),
line width ($\gamma_i$) and effective plasma frequency ($\omega_{pi})$ for the
three oxygen-related {\it c}-axis modes of YBa$_2$Cu$_3$O$_{6.95}$ ($\bullet$).
The $\approx 279$ cm$^{-1}$ mode fails to show any discontinuity at $T_c$,
while the 312 and 570 cm$^{-1}$ modes both soften below $T_c$; the 570
cm$^{-1}$ feature also appears to gain oscillator strength below $T_c$. The
rotation angle used to fit the asymmetric line shape is shown in the O(4) panel
($\circ$), and shows a slightly increasing asymmetry with decreasing
temperature.}
\vspace*{-0.5cm}%
\centerline{\includegraphics[width=16.3cm]{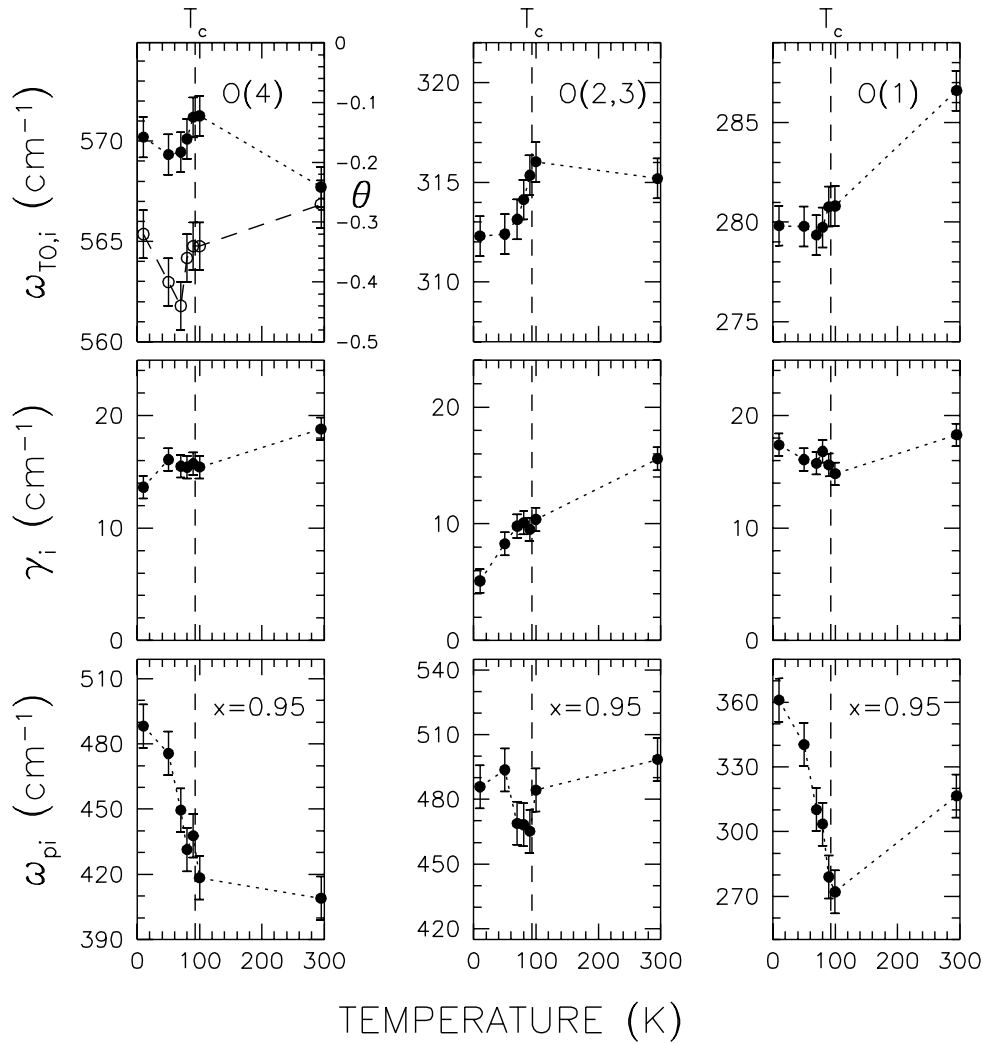}}%
\vspace*{-1.6cm}%
\label{phona}
\end{figure*}
%
%

As the doping decreases to $x\approx 0.7$, the 155~cm$^{-1}$ O(4) mode softens
and decreases in strength, while the mode at 148~cm$^{-1}$ hardens slightly,
and gains strength. Below $x\approx 0.7$, it is clear that the fundamental
nature of this mode has changed; the 155 cm$^{-1}$ feature hardens
dramatically, and loses almost all of its oscillator strength; in contrast, the
148 cm$^{-1}$ mode softens, and its oscillator strength approaches the value of
the 155 cm$^{-1}$ mode in the highly-doped material.

The 312~cm$^{-1}$ plane-oxygen mode displays drastically different behavior at
room temperature and at 10~K.  At room temperature, the mode hardens from
315~cm$^{-1}$ to 320 cm$^{-1}$ and its oscillator strength remains unchanged as
the doping is reduced.  However, at 10~K, this mode softens dramatically with
decreasing doping, furthermore, at $x\approx 0.5$ this mode, at low
temperature, has lost most of its considerable oscillator strength.  The loss
of oscillator strength at low temperature in the underdoped materials results
from the transfer of their spectral weight to the 400 cm$^{-1}$ mode, which
will be discussed in more detail below.

In the fully-oxygenated material the 610~cm$^{-1}$ mode can just be seen as a
weak side band of the 570~cm$^{-1}$ mode, but as the doping is reduced the
610~cm$^{-1}$ grows  substantially in spectral weight. The mode hardens
substantially to 645~cm$^{-1}$ in the $x=0.5$ material, while the 570~cm$^{-1}$
mode that it is associated with is steadily losing oscillator strength with
decreasing doping.  This effect has been attributed by Burns et al. to the
shortening of the O(4)-Cu(1) bond by 6\% in the same concentration range
\cite{Hazen91}.  In rough agreement with this model the 610 cm$^{-1}$ line
changes its frequency by 6\% as well in the same concentration range. In Burns'
picture the 570 cm$^{-1}$ line arising from the O(4) at the four-fold
coordinated copper, {\it softens} by 3\%, a value typical of the overall {\it
c}-axis expansion, suggesting a {\it lengthening} of that bond. If the
interpretation of Burns is correct we would expect a modulation of the O(4)
position in the {\it a}-direction by approximately 0.03~\AA\ (1~\AA\ $=
10^{-10}$~m) in the Ortho~II structure where there is an alternation between
two- and four-fold coordinated coppers. There have been reports, by several
authors \cite{Conradson90,Mustre90,Stern93}, of a double peak in the
Cu(1)--O(4) bond length distribution measured by XAFS by as much as 0.1~\AA .

%
%
\begin{figure*}[t]
\caption{The temperature dependence of the frequency ($\omega_{{\rm TO},i}$),
line width ($\gamma_i$), and effective plasma frequency ($\omega_{pi}$) for the
three oxygen-related {\it c}-axis modes of YBa$_2$Cu$_3$O$_{6.70}$, and a new
fourth feature associated with the change in coordination of the Cu(1) atom,
discussed in the text ($\bullet$).  The feature at $\approx 285$~cm$^{-1}$
softens with decreasing temperature, while the two modes at $\approx 557$
cm$^{-1}$ and $\approx 630$ cm$^{-1}$ harden smoothly; the feature at 318
cm$^{-1}$ softens well above $T_c$.  The three high-frequency modes begin to
loose oscillator strength at $\approx 150$~K, while the low-frequency mode
gains in strength primarily below $T_c$.  The rotation angles ($\circ$) used to
fit the two high-frequency modes are shown in the O(4)$^\ast$ and O(4) panels:
the $\approx 570$ cm$^{-1}$ mode becomes increasing asymmetric at low
temperature, while the mode at $\approx 610$~cm$^{-1}$ becomes more symmetric.}
\vspace*{-0.4cm}%
\centerline{\includegraphics[width=16.3cm]{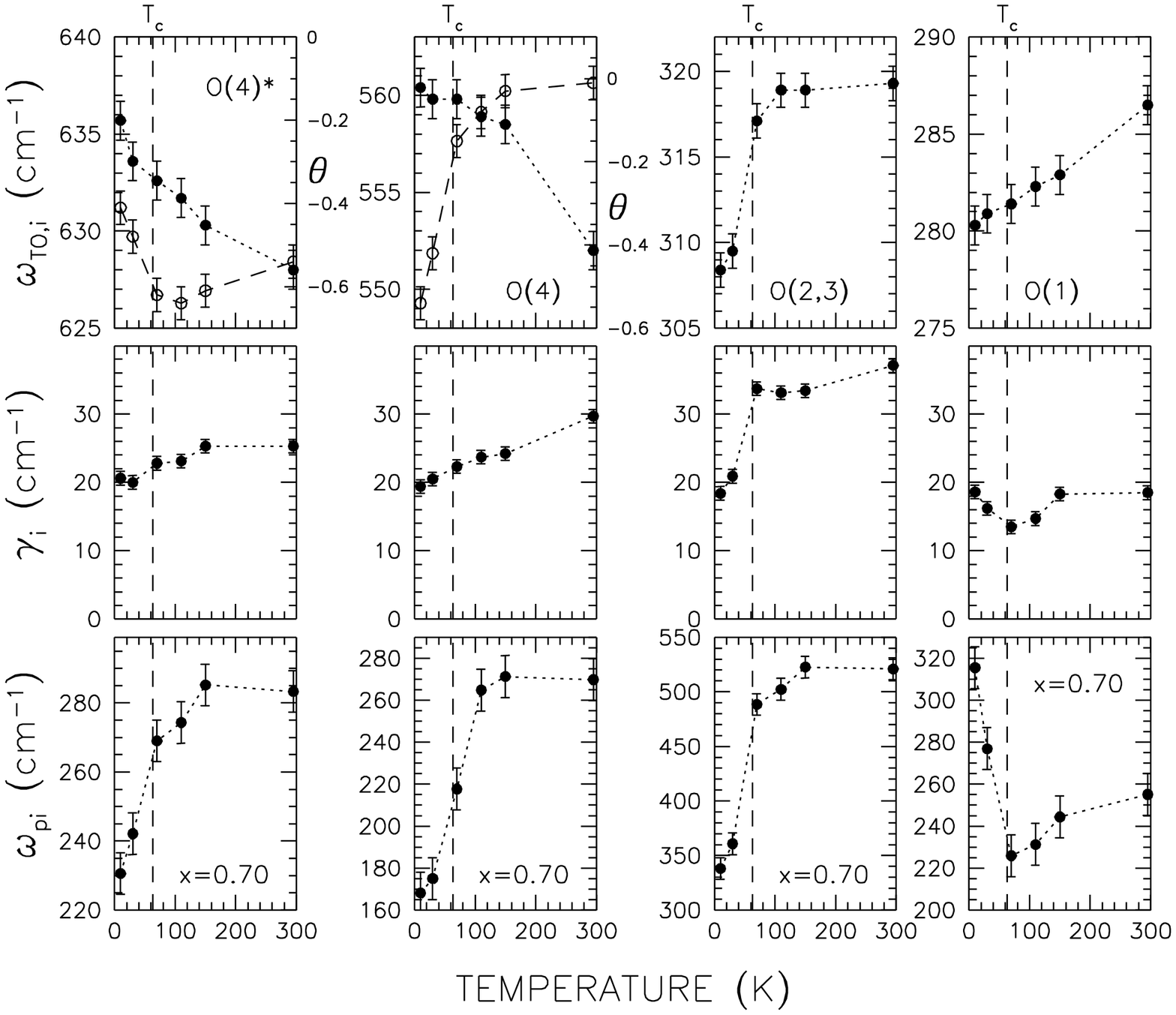}}%
\vspace*{-1.4cm}%
\label{phonb}
\end{figure*}
The widths of the phonons increase steadily with decreasing doping, with the
exception of the 155~cm$^{-1}$ mode, which shows some signs of narrowing at low
dopings.  The steadily increasing widths may be a reflection of the O(1)-site
disorder at low dopings. However, if there is some long-range ordering of the
O(1) sites at low dopings, as is expected in the Ortho~II phase, then there
should be some narrowing observed in the lines.  While there is some evidence
of this in the 155~cm$^{-1}$ mode, the 279 cm$^{-1}$ mode shows no signs of
narrowing at low-oxygen dopings.


The oscillator strengths are a convenient guide by which to follow the effects
of the removal of oxygen from the O(1) sites. In particular, this is true for
the 279~cm$^{-1}$ mode, which is almost exclusively an O(1)-oxygen vibration.
In the $x=0.5$ material, we expect that the O(1) sites should be half empty. At
295~K, the effective-plasma frequency of this mode decreases from $\approx 310$
cm$^{-1}$ to $\approx 200$ cm$^{-1}$ in the $x=0.5$ material; taking the
na\"{\i}ve view that the ratio of the oscillator strengths ($\propto
\omega_p^2$) of 0.44 should be a reflection of the half-empty occupancy of the
O(1) sites, we find good agreement.

The asymmetric line shapes of the 570 and 610~cm$^{-1}$ modes require that they
be fitted using a rotated Lorentzian. While this line shape generally satisfies
oscillator sum rules for small rotations, it has been pointed out that for
large rotations ($\theta > 0.4$ rad) this is no longer true.  The values for
$\omega_p$ may generally have increasingly large errors associated with them
for large rotations. The results show a decreasing oscillator strength for the
570 cm$^{-1}$ with decreasing doping, and the 610~cm$^{-1}$ mode show a
steadily increasing oscillator strength with decreasing doping with the two
oscillator strengths becoming equal at $x=0.7$. Below  $x\approx 0.6$, the 610
cm$^{-1}$ line appears to be decreasing in intensity. This latter result is
consistent with the small value for $\omega_p$ observed in the tetragonal
YBa$_2$Cu$_3$O$_6$ material.

%
%
\subsection{Temperature dependence of the phonon parameters}
The temperature dependence of the phonon parameters for the three
high-frequency modes in YBa$_2$Cu$_3$O$_{6.95}$ is shown in Fig.~\ref{phona},
the results are in general agreement with previous work
\cite{Bonn87b,Wittlin87a,Litvinchuk92,Macfarlane87}.  The temperature
dependence for the four high-frequency modes in YBa$_2$Cu$_3$O$_{6.7}$ is shown
in Fig.~\ref{phonb}.

In the highly-doped material, as the temperature is lowered, the phonon
frequencies generally increase due to anharmonic effects, but there are
exceptions, most notably the 279 cm$^{-1}$ chain-oxygen mode, which softens by
$\approx 6$~cm$^{-1}$ between  295 and 10~K. The two low-frequency modes
display little temperature dependence. The mode at $\approx 155$ cm$^{-1}$
hardens slightly from 153 cm$^{-1}$ at room temperature to 155 cm$^{-1}$ at
10~K.  The feature at $\approx 194$ cm$^{-1}$ softens to 193 cm$^{-1}$ at 10~K;
this feature shows a slight anomaly at $T_c$.

The three remaining modes at 279, 312 and 570 cm$^{-1}$ are all associated with
oxygen vibrations, and show a more dramatic behavior.  The 312 and
570~cm$^{-1}$ modes both harden with decreasing temperature, but show a
discontinuity and soften below $T_c$.  The peak at $\approx 279$ cm$^{-1}$,
originally associated with the oxygen atoms in the CuO$_2$ planes, has been
reassigned as an O(1) chain-oxygen vibration \cite{thomsen91}; this mode
softens continuously from room temperature to below $T_c$, and shows only a
weak anomaly at $T_c$.  However, like the 570 cm$^{-1}$ mode, the 279 cm$^{-1}$
mode shows some signs of hardening below $\approx 40$~K.  The line widths of
the 279 cm$^{-1}$ mode shows only a weak temperature dependence, while the line
widths of the 312 cm$^{-1}$ and 570~cm$^{-1}$ modes both narrow with decreasing
temperature; interestingly, none show any sign of anomalous behavior at $T_c$.

There is substantial variation in the oscillator strengths with temperature.
The 279 cm$^{-1}$ feature weakens considerably with decreasing temperature, but
then becomes much stronger below $T_c$, while the 312 cm$^{-1}$ remains
relatively constant. The O(4) mode at 570 cm$^{-1}$ shows a rapid increase in
oscillator strength at $T_c$, and initially becomes more asymmetric below
$T_c$, suggesting the possibility of electron-phonon coupling (all the other
modes have been fit using simple Lorentzian oscillators).

The sharpness of the phonons in the highly-doped phase, especially at the low
frequency modes at 155 and 279 cm$^{-1}$, where anharmonic contribution to the
width is small, is a sign of good crystal quality.  In previous single crystal
work with polished samples, these features were broad and weak, suggesting
either poor oxygen homogeneity, or that polishing the sample has induced
surface damage or altered the oxygen content of the material in the surface
layers, particularly in the chains.  Furthermore, an examination of the
reflectance of polished {\it c}-axis crystals \cite{koch90,koch90b} shows that
they bear a strong similarity to the reflectance of freshly-annealed ceramics.

The 194~cm$^{-1}$ mode, according to the shell model, has its eigenvectors
confined to plane region of the unit cell, and therefore does not split when
alternate chains become deoxygenated, as expected.

The vibrations at 279 and 312 cm$^{-1}$ in the highly-doped material harden and
soften slightly with decreasing oxygen content, respectively, while also
broadening considerably.


The temperature dependence of the phonon parameters in the oxygen-reduced
system is dramatically different from the fully-doped case.  As
Fig.~\ref{phonb} shows for YBa$_2$Cu$_3$O$_{6.70}$ there is now an additional
mode at $\approx 610$ cm$^{-1}$.  The two apical-oxygen modes at 557 and 630
cm$^{-1}$ harden continuously with decreasing temperature, and there is no sign
of a phonon ano\-ma\-ly at the superconducting transition temperature. As noted
by Lit\-vin\-chuck \cite{Litvinchuk92} the plane feature at $\approx 318$
cm$^{-1}$ displays very interesting behavior in that it begins to soften
rapidly at $\approx 100$~K, well above the $T_c$ of 63~K. The chain feature at
$\approx 285$ cm$^{-1}$ at room  temperature softens to $\approx 279$ cm$^{-1}$
at low temperature, none of these features shows any discontinuity at $T_c$.

The behavior of the line widths is similar to that of the oscillator strengths.
The strengths of the two high-frequency modes both begin to decrease at
$\approx 140$~K, well above $T_c$, as does the strength of the $\approx 312$
cm$^{-1}$ mode.  This is in sharp contrast to the $\approx 279$~cm$^{-1}$ mode,
which loses strength with decreasing temperature until $T_c$, at which point it
begins to increase very rapidly, this is the only feature which shows any
sensitivity to $T_c$.

\begin{figure}[t]
%

%
%
\caption{The optical conductivity ($\sigma_1$) of YBa$_2$Cu$_3$O$_{6.60}$ along
the {\it c}-axis as a function of temperature, from 200 to 700~cm$^{-1}$, for
four temperatures each above and below $T_c$ ($\approx 58$~K). Note the
dramatic loss in the strength of the feature at $\approx 318$ cm$^{-1}$, and
the corresponding increase in the  feature (or group of features) centered at
$\approx 400$ cm$^{-1}$, there is no anomalous behavior near $T_c$. }
\vspace*{-0.5cm}%
\hspace*{-0.4cm}%
\centerline{\includegraphics[width=10.5cm]{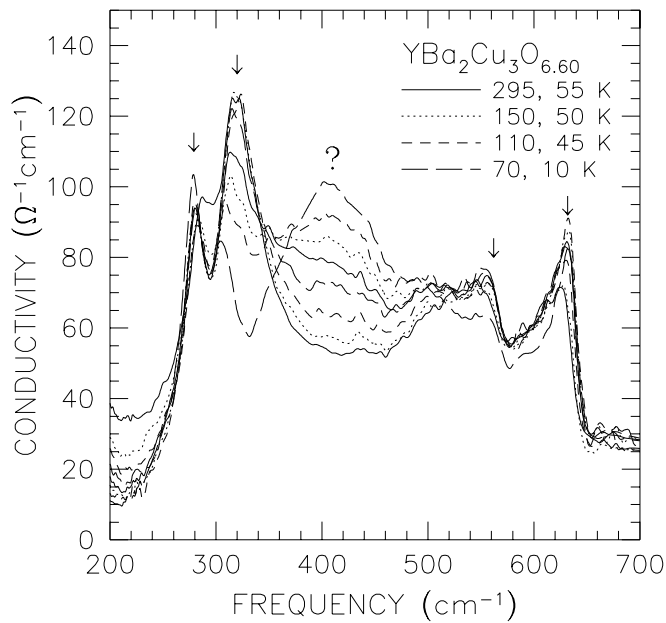}}%
\vspace*{-1.4cm}%
\label{lump}
\end{figure}

The rapid loss of oscillator strength, particularly in the 312 cm$^{-1}$ mode
in the $x=0.7$ material is even more noticeable in the $x\lesssim 0.6$
materials: in the $x=0.5$ material, this mode appears to have completely
collapsed at 10~K.  The mode at $\approx 410$ cm$^{-1}$ (or combination of
modes), which is very weak in the $x=0.95$ material, is gradually becoming
stronger as the oxygen content is decreased, so that in the oxygen-reduced
$x=0.6$ material, it is one of the dominant features.  Furthermore, as the
other oxygen related vibrations are losing oscillator strength, this feature is
becoming correspondingly stronger, as Fig.~\ref{lump} illustrates.
Interestingly, optical sum-rule calculations show no net increase or decrease
in the spectral weight in the $250-700$ cm$^{-1}$ region, suggesting that the
$\approx 400$~cm$^{-1}$ feature(s) is growing as a result of the transfer of
oscillator strength for the other modes, but in particular from the planar
copper-oxygen vibrations.  The $\approx 400$ cm$^{-1}$ feature is identifiable
well above $T_c$, and appears to grow smoothly, and without any sudden
discontinuity at $T_c$.  Furthermore, the 312 cm$^{-1}$ mode loses a great deal
of oscillator strength, and begins to do so well above $T_c$.  This suggests
that while there is a major reorganization of the vibrational energy in the
CuO$_2$ planes (in the oxygen-reduced materials) as the temperature is lowered;
this is a separate mechanism which may affect superconductivity in these
materials, but does not appear to be affected by the superconducting
transition.  Interestingly, similar behavior is observed in the high-frequency
copper-oxygen vibrations in cupric oxide (CuO) near the N\'eel transition
\cite{homes95}.

%
%
\section{Discussion}
\subsection{The assignment of the 610 cm$^{-1}$ phonon}
The most striking change at room temperature with doping, is the growth of the
610~cm$^{-1}$ mode at the  expense of the 570~cm$^{-1}$ mode.  This
610~cm$^{-1}$ feature has been observed in other work and has been the subject
of some controversy, since there is lack of agreement in the Raman community on
the presence of the corresponding Raman mode at 600 cm$^{-1}$ \cite{burns91}.
In the tetrahedral YBa$_2$Cu$_3$O$_6$ material, which lacks the O(1) oxygen
chains, there is an infrared active $A_{2u}$ mode that is observed at $645-
650$~cm$^{-1}$ which, like the 573 cm$^{-1}$ mode in the orthorhombic phase,
has been identified with the O(4) oxygen \cite{burns88}.  In removing the chain
oxygens, the four-fold coordinated Cu(1) ``squares'' become two-fold
coordinated ``sticks'' \cite{burns91}. The change in coordination number is
accompanied by a decrease in the Cu(1)--O(4) distance, and a resulting
hardening of the mode causing it to move from 570 to 610 cm$^{-1}$
\cite{burns91}. The feature at $\approx 610$ cm$^{-1}$ therefore appears to be
an $A_{2u}$ mode native to the tetrahedral phase involving the
O(4)--Cu(1)--O(4) sticks.

In accord with this picture we find that the oscillator strength of the two
modes, at 570 and 610~cm$^{-1}$, are nearly identical in the $x=0.70$ material.
One expects the ideal Ortho~II structure to occur at $x=0.50$ where there are
an equal number of two and four-fold coordinated coppers with every other chain
complete and every other chain empty.  Also, the temperature dependence of the
various mode parameters seen in Fig.~\ref{phonb}, is nearly identical for the
two modes confirming that they are located on nearly identical structural
units.

Some recent Raman work does show intensity variation with doping that parallels
our work with weakening of the 500~cm$^{-1}$ line as the 600 cm$^{-1}$ line
grows. The two lines reach lines of equal strength at 500 and 600 cm$^{-1}$ in
the oxygen reduced, $x=0.75$ material.\footnote{M.N. Iliev. Private
communication.} Surprisingly, this is observed for the \boldmath $E\parallel b$
\unboldmath polarization.

Neutron measurements, too, find a mode in oxygen-reduced material $x=0.25$ at
\boldmath $q=0$ \unboldmath at the frequency that is consistent with Burns'
model of a high frequency mode for two-fold coordinated Cu(1).\footnote{N.
Pyka. Unpublished results quoted in ref. \citen{Pintschovius94}.} It should
also be noted that the neutron density of states determined by incoherent
scattering shows a high frequency band of modes in the $x=0$ phase centered at
650 cm$^{-1}$, whereas the fully-doped material has its highest branch at
$\approx 575$ cm$^{-1}$ \cite{Renker88}, in accord with the notion of a set of
high-frequency modes associated with the oxygen-reduced material.

\subsection{The 400~cm$^{-1}$ mode}
%
%
Another trend with decreasing doping is the appearance and growth of a broad
band at $\approx 400$ cm$^{-1}$ at low temperature. The band also shifts to
lower frequencies as the doping is reduced. As shown in Fig.~\ref{sigma10} and
in Fig.~\ref{lump}, this unusual feature increases in strength in the normal
state as the temperature is lowered, but does not show anomalous behavior at
$T_c$.  At very low dopings, ($x \lesssim 0.6$) where this band is quite
strong, it appears to gain oscillator strength at the expense of the 312, 570,
and 610 cm$^{-1}$ phonons. At room temperature where the phonons are sharp and
well defined, the total spectral weight of the in the $250 - 700$~cm$^{-1}$
range, is equal to the total spectral weight, including the 400 cm$^{-1}$ mode,
at low temperature.  This conservation of spectral weight is consistent with
the idea that the broad peak at $\approx 400$ cm$^{-1}$ is indeed a phonon.

The broad, weak mode at $\approx 408$ cm$^{-1}$ has been observed in ceramic
spectra of samples that appear to have been oxygen reduced \cite{Bonn87b}, but
this mode does not correspond to any of the calculated $B_{1u}$ normal modes
\cite{bates87,kress88}.  A similar mode is seen in YBa$_2$Cu$_4$O$_8$
\cite{basov94c} as well as in Pb$_2$Sr$_2$RCu$_3$O$_8$ \cite{reedyk94}. The
common element in the three compounds is the third copper layer with two-fold
coordinated oxygen bonds. In all three systems the mode appears at low
temperature, growing in strength at the expense of both in-plane mode at 312
cm$^{-1}$ as well as the two apical-oxygen modes.  This behavior suggests that
an unusual transition is taking place that involves several atoms in the unit
cell.  The transition may be related to the  zone-boundary mode reported on by
Reichardt et al. \cite{reichardt89}; located at approximately 400 cm$^{-1}$, it
was found to be anomalously broad and involved the ``breathing'' motions of the
ions around the planar copper.  This mode is not normally infrared active, but
could become so as a result of some symmetry-breaking process.  The mode
involves the {\it c}-axis motion of the O(4), but not the in-plane motion of
the planar oxygens.  Its invocation would explain only half of our
observations, namely the loss of O(4) spectral weight to the new mode through
an electron-phonon process of the type discussed in connection with organic
molecules by Rice \cite{Rice80}. In such a process a symmetry-breaking
transformation allows totally-symmetric vibrations (here the zone-boundary
breathing mode) to become optically active.  A necessary ingredient of the
process is the presence of low-lying charge-transfer bands. The excess spectral
weight and the asymmetric shape of the O(4) bands suggests that charge is being
pumped at the O(4) frequency between chains and planes. This process is already
active at high temperature above the formation temperature of the pseudogap. It
would then not be unreasonable to anticipate that when the symmetry-breaking
transition sets in at 150 K in the oxygen-reduced materials, that some of this
phonon-driven charge is transferred from the O(4) and plane modes, to the newly
activated breathing mode.  The difficulty with this argument is that it is does
not apply to the other two compounds where this mode is observed
\cite{basov94c,reedyk94}.

Possibly related to this unusual phonon, polarized in the c-direction, is the
phenomenon in the {\it ab}-plane conductivity reported on by Reedyk and Timusk
where, in YBa$_2$Cu$_3$O$_{6.95}$ the {\it c}-axis longitudinal modes
interacted strongly with the {\it ab}-plane conductivity
\cite{timusk91,reedyk92e}. The strongest of these was a broad phonon with the
longitudinal frequency of 440 cm$^{-1}$.

The structural transition giving rise to the phonon at $\approx 400$ cm$^{-1}$
band appears to  be related to the pseudogap seen at $\approx 280$ cm$^{-1}$ in
underdoped materials.  It is strong in underdoped materials where the pseudogap
is also well developed. In Zn-doped YBa$_2$Cu$_4$O$_8$, both the pseudogap and
the 400~cm$^{-1}$ mode disappear at a 1.7\% doping level.\footnote{D.N.~Basov,
T.~Timusk and B.~Dabrowski (unpublished)} However, there are some differences.
Whereas the pseudogap appears gradually, in case of the $x=0.6$ material at a
temperature well above 300~K, the 400~cm$^{-1}$ band forms quite suddenly at
$\approx 150$~K, as shown by the sudden loss of O(4) spectral weight below this
temperature.

In summary, at this stage we have only the vaguest understanding of the nature
of the phonons giving rise to the band at 400~cm$^{-1}$, it is a feature that
appears at low temperature in oxygen-reduced materials only, and gets its
oscillator strength from plane-buckling and the O(4) modes. Its frequency
coincides with the anomalous broad mode seen in neutron scattering, and a broad
minimum in the {\it ab}-plane conductivity. Zn doping of YBa$_2$Cu$_4$O$_{8}$
seems to interfere with the formation of this mode. It should be noted that it
is not the result of a simple structural transition, since the {\it
frequencies} of the modes involved change little; there is no soft-mode
behavior.  It appears that the transition is electronic in nature, the main
result being the shift of effective charges of the phonons, not their
frequencies.

%
%
\section{Conclusions}
The phonons in the oxygen-reduced systems provide a richly detailed structure.
At high dopings, there are five strong $B_{1u}$ modes. At low dopings
($x\lesssim 0.7$), many of the lines split into a number of different
components.  In particular, the high-frequency apical-oxygen mode splits into
two components; the low-frequency component is associated with the four-fold
coordinated copper atom, and the high-frequency component with the two-fold
coordinated copper.  What is most unusual is the development of a broad
feature, centered at $\approx 400$~cm$^{-1}$.  This feature, attributed to a
(unknown) phonon, becomes more pronounced as oxygen is removed from the system;
it grows in strength as the temperature is lowered, appearing to draw
oscillator strength from the vibrations at $\approx 320$, 570 and 610~cm$^{-1}$
with decreasing temperature. This behavior develops well above $T_c$, and other
phonon anomalies are seen in the $x=0.7$ material at $\approx 150$~K.

Since these massive shifts of spectral weight to the 400~cm$^{-1}$ mode are not
seen in other high-temperature superconductors, such as La$_{2-x}$Sr$_x$CuO$_4$
and Bi$_2$Sr$_2$CaCu$_2$O$_{8+ z}$ \cite{Tamasaku94,Tajima94}, they may be
related to some special structural property of materials with a third copper
layer in the charge reservoir area of the unit cell, and not a fundamental
property of the cuprates.

\section*{Acknowledgements}
We would like to thank A.~Bianconi, J.C.~Irwin, W.~Reichardt, and J.~R\"ohler
for stimulating discussions.  This work was supported by the Natural Sciences
and Engineering Research Council of Canada and the Canadian Institute for
Advanced Research.

%
%

\BalanceColumns[0.5]

%
%
%
%
%
\end{document}